\documentclass[journal]{IEEEtran}
\usepackage{epsfig,graphicx,rotating}
\usepackage{color}
\usepackage{tabularx}
\usepackage{multicol,multirow,longtable}
\usepackage{amssymb,amsmath}
\usepackage{epstopdf}
\usepackage{placeins}
\usepackage{color}
\usepackage{bbm}
\usepackage{amsfonts}
\usepackage{subcaption}
\usepackage{cite}
\usepackage{mathrsfs}
\usepackage[T1]{fontenc}
\usepackage[utf8]{inputenc}
\usepackage{booktabs}
\usepackage{fix-cm}
\begin{document}
\title{Spectral Efficiency Analysis in Presence of Correlated Gamma-Lognormal Desired and Interfering Signals}
\author{Aritra Chatterjee,~\IEEEmembership{Student Member,~IEEE,}
Sandeep Mukherjee,~\IEEEmembership{Member,~IEEE,}        
         and Suvra Sekhar Das,~\IEEEmembership{Member,~IEEE}
\thanks{The authors are with the G. S. Sanyal School of Telecommunications, Indian Institute of Technology Kharagpur, India (e-mail: aritrachatterjee@iitkgp.ac.in; sandeep@gssst.iitkgp.ernet.in; suvra@gssst.iitkgp.ernet.in). Corresponding author: A. Chatterjee.}         
}
\maketitle
\begin{abstract}
  Spectral efficiency analysis in presence of correlated interfering signals is very important in modern generation wireless networks where there is aggressive frequency reuse with a dense deployment of access points. However, most works available in literature either address the effect of correlated interfering signals or include interferer activity, but not both. Further, available literature has also addressed the effect of large scale fading (shadowing and distance-dependent path loss) only, however has fallen short of including the composite effect of line of sight and non-line of sight multipath small scale fading. The correlation of desired signals with interfering signals due to shadowing has also not been considered in existing literature. 
  In this work, we present a comprehensive analytical signal to interference power ratio evaluation framework addressing all the above mentioned important components of the model in a holistic manner. In this analysis we extend and apply the Moment Generating Function-matching method to such systems so that correlation and activity of lognormal random variables can be included with high accuracy.
  We compare the analytical results against realistic channel model based extensive Monte-Carlo simulation for mmWave and sub-6 GHz in both indoor and outdoor scenarios. 
  The performance of the model is depicted in terms of mean, $\alpha\%$ outage spectral efficiency and Kullback–Leibler divergence and Kolmogorov–Smirnov distance.
\end{abstract}\vspace{-0.35cm}
\section{Introduction}
To meet the increasing demand for high throughput connectivity, modern wireless networks are adopting methods of aggressive frequency reuse with dense deployment of access points (AP)/ base stations (BS) \cite{Andrews_2014}. 
Simulation based performance analysis of such interference limited radio access networks (RAN) is highly complex and time consuming which motivates the development of analytical methods \cite{Cardieri_2001}.\\  
\indent In order to develop realistic analytical performance evaluation for above mentioned RAN scenario, one has to consider the holistic effect of line of sight (LOS) and non-line of sight (NLOS) multipath small scale fading (SSF), large scale fading (LSF) including both shadowing and distance-dependent path loss (PL), interferer activity and  shadowing correlation between received signals as well as between desired and aggregate interfering signals \cite{3gpp36814}. In the following paragraph, we discuss the literature available on performance analysis of RAN in presence of co-channel interference (CCI). \\
\indent Earlier works on CCI-based performance analysis \cite{Abu_Dayya_1991, Alouni_1999} do not consider interferer activity or correlation among interfering signals. Although in \cite{Tellambura_1999} expression of outage probability in CCI condition is given, yet it does not account for LSF. In \cite{abudayya_1994} correlated interferers have been considered, but the effect of multipath SSF and interferer activity has been neglected. In \cite{fischione_2007}, the activity of neighboring interferers is modeled as Bernoulli random variables (RV), however the effect of correlated interfering signal has not been taken into account. SSF as well as LSF (path loss and shadowing) along with interferer activity is considered in \cite{Chandhar_2014} but shadowing correlation among desired and interfering signals have been overlooked, which is present in realistic propagation conditions \cite{3gpp36814}.
Therefore we find that the comprehensive effects of LOS or NLOS SSF, shadowing, distance-dependent path loss, interferer activity and shadowing correlation have not been considered in unified manner in existing system models, which is necessary for a holistic performance analysis. \\
\indent The CCI appears at the receiver as sum of correlated lognormal (LN) distributed RVs. There exists no known closed form expression for the distribution of sum of LN RVs \cite{stueber_2001}. Therefore we now look into the approximation methods available in literature. Popular methods to be used are `Fenton-Wilkinson's (F-W)' \cite{fenton_1960} and `Schwartz and Yeh’s (S-Y)' \cite{schwartz_1982}. These methods use the moment-matching (MM) approach. The works \cite{abudayya_1994, fischione_2007, Chandhar_2014} described earlier use MM methods but they are known to be useful for standard deviation of LN RVs ($\sigma_{\chi}$) within $\sim 4$ dB only. 
A variant of F-W method, namely `truncated lognormal approximation', proposed in  \cite{Chatterjee_2018} is valid for a wider range of $\sigma_{\chi}$ ($4-12$ dB). But, the approach is usable for uncorrelated RVs only. 
The moment generating function (MGF) matching approach \cite{mehta_2007} has been shown to be of higher accuracy over a wider range of $\sigma_{\chi}$, albeit at the cost of increased computational complexity. In \cite{chandhar_2016}, MGF matching method has been extended to include Bernoulli RV but not correlation of LN RVs. Therefore, we find that on one hand we have MM methods which are easy to compute but suffers from low accuracy and have limited applicable range of parameter values, on the other hand we have MGF-matching method which provides higher accuracy but is still incomplete for use.   \\
\indent Therefore to summarize, we find that the system models in existing literature on analytical CCI evaluation falls short of capturing the comprehensive realistic co-channel interference scenario accurately which is addressed in this work by extending the MGF matching method to suitably include both correlation and activity to analyze the holistic effects of SSF (LOS/NLOS), LN distributed shadowing, distance-dependent PL, interferer activity and shadowing correlation between desired and interfering signals.\\
\indent Since fifth generation (5G) mobile networks are destined to use millimeter wave (mmWave) spectrum \cite{Andrews_2014, Rappaport_2013, Sun_2018}, we present the downlink signal to interference power ratio (SIR) based spectral efficiency (SE) evaluation of a recently conducted measurement campaign reported in \cite{maccartney_2015}. We also show the effectiveness of our derived expressions in sub-6 GHz band\footnote{In this work, by the term `sub-6 GHz frequency band' we mean carrier frequency ($f_c$) of 2 GHz, which is considered as one of the major frequency bands of operation in fourth-generation long term evolution (4G-LTE) and beyond \cite{3gpp36873}} which would still remain an important part of 5G, especially for outdoor coverage. 
The analytical framework presented here is shown to remain useful for realistic values of system parameters  as outlined in \cite{maccartney_2015} for mmWave scenarios and in existing 3GPP prescribed channel models \cite{3gpp36873} which are suggested to be used for evaluation of 5G and beyond systems for sub-6 GHz.  
The main contributions of this work are:
\begin{itemize}
\item A holistic system model representing a co-channel RAN has been developed to include the realistic propagation effects of SSF (LOS/NLOS), shadowing, distance-dependent PL, interferer activity and shadowing correlation.  
\item An MGF-based approximation has been extended for the distribution of sum of correlated Bernoulli-LN RVs in order to incorporate both interferer activity and correlation. 
\item{An exact expression of correlation coefficient between the power of the desired AP and the CCI has been derived.}
\item{Applying the above results in the developed system model, we have presented an extensive performance analysis of a downlink co-channel indoor as well as urban macro scenario operating in mmWave as well sub-6 GHz band.}
 \end{itemize}

%
\vspace{-0.3cm}
\section{System Model and Assumptions}
\label{sec:system_model}
In this work, the downlink of a typical co-channel RAN with a total of $K+1$ AP has been considered. 
User equipments (UE) situated in any location is served by the nearest AP. Thus, for any UE, there is one desired AP, remaining $K$ APs act as interferers.
We consider single input single output configuration. Consideration of multiple antennas in APs, UEs or both can be seen as a potential future work. All the APs and UEs use omnidirectional antennas with vertical polarization. The transmit power of the APs is denoted by $P_T$. We consider a block fading channel model where all Tx-Rx links are affected by both large and small scale fading, which are modeled by LN and Gamma distributions respectively \cite{stueber_2001}, \cite{3gpp36873}. 
We consider an interference-limited scenario, where the received interference power dominates the noise power. 
The activity of the $k^{th}$ interfering AP is modeled by $\nu_k$, which is a Bernoulli RV with probability mass function:  $\mathbb{P}(\nu_k = 1) = p_k$ and $\mathbb{P}(\nu_k = 0) = 1-p_k, ~ 0\leq p_k\leq 1$. When there exists no UE under the coverage region of a particular AP, it is said to enter `inactive' state. The specific value of $p_k$ depends on the traffic model and load analysis on that particular AP, calculation of which is beyond the scope of this work. A detailed discussion on this topic can be found in \cite[Sec. II.A]{Chandhar_2014}, \cite{pokhariyal_2007}, \cite{barbarossa_2011}.   
We also assume that $\left\lbrace \nu_1,\nu_2,\hdots,\nu_K \right\rbrace$ are uncorrelated. 
The combined received power by a typical UE from desired and interfering APs can be expressed as,
\begin{align}
P_R &= P_T \times \left[ 10^{-0.1\mathscr{L}(d_0)}\chi_0|h_0|^2 \right. \nonumber \\
& \left. + \sum_{k=1}^K \left( \nu_k10^{-0.1\mathscr{L}(d_k)}\chi_k|h_k|^2\right)\right],
\label{eq:total_power}
\end{align}
where the first term denotes the power received from the desired AP and the second term denotes the aggregate power from all the interfering APs. The distance from the UE to the serving AP is denoted by $d_0$, whereas the distance to the $k$-th interferer is $d_k$. The interference power from different interferers are considered to be correlated, due to the proximity of the interfering APs.
In (\ref{eq:total_power}), $\mathscr{L}(d)$ denotes the average path loss (in dB) for a Tx-Rx distance of $d$ for LOS or NLOS link, depending on the nature of the link, as detailed in \eqref{eq:PL_LOS_NLOS_mmW}.
\begin{equation}
\mathscr{L}(d) = \overline{PL}(d_{ref}) + 10n_{L/NL}log(d/d_{ref}),
\label{eq:PL_LOS_NLOS_mmW}
\end{equation}
where, $n_{L}$ and $n_{NL}$ are the path loss exponents (PLE) for LOS and NLOS cases respectively and $\overline{PL}(d_{ref})$ is the pathloss at a reference Tx-Rx distance ($d_{ref}$). 
The effect of shadowing is captured by $\chi$ such that $10log_{10}(\chi) \sim \mathcal{N}\left( 0,\sigma_\chi^2\right)$, where $\sigma_\chi$ denotes shadowing standard daviation, whose values in mmW and sub-6 GHz for LOS and NLOS have been provided in \cite[Table 10]{maccartney_2015} and \cite[Table 7.2-1]{3gpp36873} respectively. The effect of small scale fading is captured in $h_k$, such that $|h_k|^2$ is Gamma distributed with shape factor of $m$ and scale factor of $\frac{\mathscr{L}(d)\chi}{m}$. In case of NLOS links, the value of $m$ equals $1$. In LOS links, the value of $m$ is related to the corresponding Rician $K$-parameter value.  
The SIR (denoted as $\Gamma$) experienced by the typical UE can be obtained from desired and interference terms in \eqref{eq:total_power},
\begin{align}
\Gamma &= \dfrac{10^{-0.1\mathscr{L}(d_0)}\chi_0|h_0|^2}{\sum_{k=1}^K \left( \nu_k10^{-0.1\mathscr{L}(d_k)}\chi_k|h_k|^2\right)} \nonumber \\
&\overset{\left( a \right) }{\approx} \frac{10^{-0.1\hat{\mathscr{L}}(d_0)}\hat{\chi_0}}{\sum_{k=1}^K \left( \nu_k10^{-0.1\hat{\mathscr{L}}(d_k)}\hat{\chi_k}\right)},
\label{eq:SIR}
\end{align}
where the approximation in step $(a)$ in \eqref{eq:SIR} is due to the fact that a product of Gamma and LN RVs can be approximated by a LN variable (denoted by $\hat{\chi_k}$) with slight modification of its mean and variance. The modified shadowing standard deviation is denoted by $\hat{\sigma_\chi}$. The detailed steps for this approximation can be found in \cite[Sec. 4.2.1]{stueber_2001} and thus have been omitted here for brevity. 
In order to find the composite distribution of $\Gamma$, it is necessary to find the distribution of the CCI (the denominator in \eqref{eq:SIR}), which has been expressed as a sum of correlated Bernoulli-LN RVs. The exact expression of the distribution of such a random variable can not be derived. In this work, we approximate the distribution of aggregate CCI with another lognormal, whose parameters are obtained using MGF-matching, as detailed in Sec. \ref{MGF_deriv}.  
\vspace{-0.5cm}
\section{Distribution of downlink SIR and SE}\label{MGF_deriv}
The downlink SIR experienced by the UE in \eqref{eq:SIR} can be rewritten as: 
\begin{align}
\Gamma =  \frac{10^{-0.1\hat{\mathscr{L}}(d_0)}\hat{\chi_0}}{\sum_{k=1}^K \left( \nu_k10^{-0.1\hat{\mathscr{L}}(d_k)}\hat{\chi_k}\right)} &= \frac{10^{0.1X_0}}{\sum_{k=1}^K \nu_k 10^{0.1X_k}} \nonumber \\
&= \frac{Y_0}{\sum_{k=1}^K \nu_k Y_k}.
\label{eq:SIR_simplified}
\end{align}
The interfering powers from $K$ interfering APs (in dB domain) are represented as correlated and not necessarily identical RVs $\lbrace X_1,\hdots,X_k,\hdots,X_K\rbrace$, where, $X_k \sim \mathcal{N}(\mu_{X_k},\sigma_{X_k}^2)$, $\mu_{X_k}$ and $\sigma_{X_k}$ denote the modified values of negative of $\overline{PL}$ (denoted as $\overline{\hat{PL}}$) and $\sigma_\chi$ (denoted as $\hat{\chi}_k$) respectively for the link between the $k^{th}$ interfering AP and UE. We define corresponding powers in linear domain as
\begin{equation}
Y_k = 10^{0.1X_k} = exp(\frac{X_k}{\xi}),~~~~ \text{where}, \xi = \frac{10}{ln(10)}.
\label{eq:X_and_Y}
\end{equation}
We define a new LN distributed RV, $Y$, such that,
\begin{equation}
Y \approx  \sum_{k=1}^K \nu_kY_k.
\label{eq:Y}
\end{equation}
Our aim in this section is to obtain the mean ($\mu_X$) and standard deviation ($\sigma_X$) of the normal distributed RV $X$ by matching the MGF of left and right hand side of \eqref{eq:Y} such that,
\begin{equation}
X = 10log_{10}(Y).
\label{eq:X}
\end{equation}
The MGF of $\sum_{k=1}^K \nu_k Y_k$ is defined as,
\begin{align}
\label{eq:mgf_exp}
&\Psi_{\left( \sum_{k=1}^K \nu_kY_k\right) } (s)  \nonumber \\
&=\mathbb{E}_{\left\lbrace \nu_1,Y_1,\nu_2,Y_2,\hdots,\nu_K,Y_K\right\rbrace}\left[ exp\left( -s\sum_{k=1}^K \nu_k Y_k\right) \right] \nonumber \\
&= \mathbb{E}_{\left\lbrace \nu_1,X_1,\nu_2,X_2,\hdots,\nu_K,X_K\right\rbrace }\left[exp\left(-s\sum_{k=1}^K \nu_kexp\left(\dfrac{X_k}{\xi}\right)\right)\right].
\end{align}
\begin{figure*}
\small
\begin{align}
&\Psi_{\left( \sum_{k=1}^K \nu_kY_k\right) }(s)
=\mathbb{E}_{\left\lbrace X_1,\hdots,X_K \right\rbrace}\left[\mathbb{E}_{\left\lbrace \nu_1,\hdots,\nu_K \right\rbrace}\left[ exp\left(-s\sum_{k=1}^K \nu_kexp\left(\dfrac{X_k}{\xi}\right)\right)\right]  \right] 
&\overset{\left( b \right) }{=}\mathbb{E}_{\left\lbrace X_1,\hdots,X_K \right\rbrace}\left[\prod_{k=1}^K \left( 1 - p_k + p_kexp\left(-s \times exp\left(\dfrac{X_k}{\xi}\right)\right)\right) \right].
\label{eq:separate_x_nu}
\end{align} \\
\begin{equation}
\Psi_{\left( \sum_{k=1}^K \nu_kY_k\right) }(s) = \int_{z_1=-\infty}^{\infty}\hdots\int_{z_K=-\infty}^{\infty} \prod_{k=1}^K \left( 1 - p_k + p_k exp\left(-s \times exp\left(\dfrac{\sqrt{2}}{\xi}\sum_{j=1}^Kc'_{kj}z_j + \dfrac{\mu_k}{\xi}\right)\right)\right)\dfrac{1}{(\pi)^{(K/2)}}exp\left( -\mathbf{z}^H\mathbf{z}\right)dz_1dz_2\hdots dz_K
\label{eq:transform}
\end{equation}
\rule{0.97\textwidth}{0.5pt}
\vspace{-0.5cm}
\end{figure*}
\normalsize
Since, $\forall k$, $X_k$ and $\nu_k$ are independent in \eqref{eq:mgf_exp}, we can separate $X_k$ and $\nu_k$ as done in (\ref{eq:separate_x_nu}) shown at the top of the next page. Step $(b)$ in (\ref{eq:separate_x_nu}) is due to the independence of $\left\lbrace \nu_1,\nu_2,\hdots,\nu_K \right\rbrace$. The expectation in final expession of \eqref{eq:separate_x_nu} has to be taken over the joint distribuion of $K$ correlated Gaussian RVs as given below:
\begin{equation}
\nonumber
p_{\mathbf{x}}(\mathbf{x}) = \dfrac{1}{(2\pi)^{(K/2)}|\mathbf{C}|^{\frac{1}{2}}}exp\left( -\dfrac{((\mathbf{x}-\boldsymbol{\mu})^H\mathbf{C}^{-1}(\mathbf{x}-\boldsymbol{\mu}))}{2}\right) ,
\label{eq:joint_distribution}
\end{equation}
where, $\mathbf{x} = \left[ X_1,X_2,\hdots,X_K\right]^T$, $\boldsymbol{\mu} = \mathbb{E}_{\mathbf{x}}\left[ \mathbf{x}\right]  = \left[ \mu_{X_1},\mu_{X_2},\hdots,\mu_{X_K}\right]^T$ and $\mathbf{C} = \mathbb{E}_{\mathbf{x}}\left[ (\mathbf{x}-\boldsymbol{\mu}){(\mathbf{x}-\boldsymbol{\mu})}^H\right]$. Using the following transformation in \eqref{eq:separate_x_nu}, we obtain \eqref{eq:transform} shown in the next page.
\begin{equation}
\label{eq:transform_defn}
\mathbf{x}=\sqrt{2}\mathbf{C}_{sq}\mathbf{z} + \boldsymbol{\mu},
\end{equation}
where, $\mathbf{C}_{sq}\mathbf{C}_{sq}^H=\mathbf{C}$. From \eqref{eq:transform_defn}, it can be written that 
\begin{equation}
x_k = \sqrt{2}\sum_{j=1}^K c'_{kj}z_j + \mu_k,
\end{equation}
where, $c'_{kj}$ is the $\left(k,j\right)^{th}$ element of $\mathbf{C}_{sq}$, $x_k$ and $\mu_k$ are the $k^{th}$ element of $\mathbf{x}$ and $\boldsymbol{\mu}$ respectively and $z_j$ is the $j^{th}$ element of $\mathbf{z}$. Taking Gauss-Hermite expansion in (\ref{eq:transform}) with respect to $z_1$, we obtain (\ref{eq:ghe_1}), where, $R_N^{(1)}$ is the remainder term, which can be neglected if the number of summands ($N$) are taken sufficiently high. The weights and abscissas of Gauss-Hermite polynomial are denoted by $w_n$ and $a_n$ respectively, which can be obtained from \cite[Table 25.10]{abramowitz_1972}. Proceeding in the same manner and taking Gauss-Hermite expansion with respect to $z_2$, $z_3$, $\hdots$, $z_N$ and subsequently dropping the remainder term, we finally express the MGF of R.H.S. of \eqref{eq:Y} by \eqref{eq:gh_approx} shown at the top of the next page.
\begin{figure*}
\small
\begin{align}
&\Psi_{\left( \sum_{k=1}^K \nu_kY_k\right) }(s) = \int_{z_2=-\infty}^{\infty}\hdots\int_{z_K=-\infty}^{\infty}  \dfrac{1}{\pi^{(K-1)/2}} exp\left(-\sum_{i=2}^K |z_i|^2\right)\sum_{n_1=1}^N \dfrac{w_{n_1}}{\sqrt{\pi}} \nonumber \\ 
&\times\prod_{k=1}^K \left( 1 - p_k + p_k exp\left(-s \times exp\left(\dfrac{\sqrt{2}}{\xi}\sum_{j=2}^Kc'_{kj}z_j + \dfrac{\sqrt{2}}{\xi}c'_{k1}a_{n_1} + \dfrac{\mu_k}{\xi}\right)\right)\right) dz_2dz_3\hdots dz_K + R_N^{(1)}
\label{eq:ghe_1}
\end{align} \\
\begin{equation}
\hat{\Psi}_{\left( \sum_{k=1}^K \nu_kY_k\right) }(s;\boldsymbol{\mu},\mathbf{C}) \approx \sum_{n_1=1}^N \hdots \sum_{n_K=1}^N \dfrac{w_{n_1}w_{n_2}\hdots w_{n_K}}{\pi^{(K/2)}} \prod_{k=1}^K \left( 1 - p_k + p_k exp \left(  -s\left[ exp \left( \dfrac{\sqrt{2}}{\xi} \sum_{l=1}^K c'_{kl}a_{n_l} + \dfrac{\mu_k}{\xi}\right) \right]\right) \right)
\label{eq:gh_approx}
\end{equation}
\rule{0.97\textwidth}{0.5pt}
\vspace{-0.5cm}
\end{figure*}
\normalsize
In a similar fashion, the MGF of L.H.S. of \eqref{eq:Y} can be expressed as: 
\begin{equation}
\nonumber
\hat{\Psi}_Y(s;\mu_X,\sigma_X) \approx \sum_{n=1}^N \dfrac{w_n}{\sqrt{\pi}}exp\left[-s\times exp\left(\dfrac{\sqrt{2}\sigma_Xa_n+\mu_X}{\xi}\right)\right].
\end{equation}
Finally, we obtain $\mu_X$ and $\sigma_X$ by solving the following pair of equations for $i = 1,2$.
\begin{equation}
\hat{\Psi}_Y(s_i;\mu_X,\sigma_X) = \hat{\Psi}_{\left( \sum_{k=1}^K \nu_kY_k\right) }(s_i;\boldsymbol{\mu},\mathbf{C}),
\label{eq:final}
\end{equation}
For the choice of suitable values of $s_i$ for accurate approximation, one is referred to \cite[Sec. VI]{mehta_2007}.
\vspace{-0.5cm}
\subsection{Correlation coefficient between the desired signal and the sum of interferers}
In this section we obtain the correlation between $X_0$ and $X$ defined in \eqref{eq:SIR_simplified} and \eqref{eq:X} respectively, denoted by $\rho_{X_0 X}$. It can be derived from \eqref{eq:Y} that the standard daviation of $Y_k$ can be expressed as 
\begin{equation}
\nonumber
\sigma_{Y_k} = \sqrt{\left[ exp\left( \left( \dfrac{\sigma_{X_k}}{\xi}\right) ^2\right) -1 \right] exp\left( \dfrac{2\mu_{X_k}}{\xi} + \left(\dfrac{\sigma_{X_k}}{\xi}\right)^2 \right)}.
\label{eq:sigma_Y}
\end{equation}  
The correlation between $Y_0$ and $Y$ has been derived in \eqref{eq:rho_Y0Y}, where the step (c) follows due to the independence of $\nu_k$ and $Y_k$. Finally, $\rho_{X_0,X}$ can be directly obtained by combining \eqref{eq:rho_Y0Y} and \eqref{eq:rho_N2LN} shown in the next page. The detailed proof of \eqref{eq:rho_N2LN} has been provided in the accompanying supplementary sheet. 
\begin{figure*}
\small
\begin{align}
&\rho_{Y_0Y} = \dfrac{\mathbb{E}_{\lbrace Y_0,Y\rbrace }\left[Y_0 Y \right] - \mathbb{E}_{Y_0}\left[ Y_0 \right] \mathbb{E}_Y\left[ Y \right]}{\sigma_{Y_0}\sigma_Y} \nonumber \\
&\approx \dfrac{\mathbb{E}_{\lbrace Y_0,\nu_1,Y_1,\nu_2,Y_2,\hdots,\nu_K,Y_K\rbrace }\left[Y_0 \left( \sum_{k=1}^K \nu_k Y_k \right) \right] - \mathbb{E}_{Y_0}\left[ Y_0 \right] \sum_{k=1}^K \mathbb{E}_{\lbrace \nu_k,Y_k\rbrace}\left[\nu_k Y_k \right]}{\sigma_{Y_0}\sigma_Y}\nonumber \\
&\overset{\left( c \right) }{=}\dfrac{ \sum_{k=1}^K \left(\mathbb{E}_{\lbrace Y_0,Y_k\rbrace }\left[  Y_0 Y_k  \right]\mathbb{E}_{\nu_k}\left[ \nu_k \right]\right) - \mathbb{E}_{Y_0}\left[ Y_0 \right] \sum_{k=1}^K \left( \mathbb{E}_{\lbrace Y_k\rbrace}\left[Y_k \right]\mathbb{E}_{{\nu_k}}\left[ \nu_k\right]\right)  }{\sigma_{Y_0}\sigma_Y} \nonumber \\
&=\dfrac{ \sum_{k=1}^K p_k\left( \mathbb{E}_{\lbrace Y_0,Y_k\rbrace }\left[  Y_0 Y_k  \right]-\mathbb{E}_{Y_0}\left[ Y_0 \right] \mathbb{E}_{\lbrace Y_k\rbrace}\left[ Y_k \right]\right)}{\sigma_{Y_0}\sigma_Y} = \sum_{k=1}^K \dfrac{p_k \rho_{Y_0,Y_k}\sigma_{Y_0}\sigma_{Y_k}}{\sigma_{Y_0}\sigma_Y} = \sum_{k=1}^K \dfrac{p_k \rho_{Y_0,Y_k}\sigma_{Y_k}}{\sigma_Y}
\label{eq:rho_Y0Y}
\end{align}
\begin{equation}
\rho_{X_0,X} = \xi^2  \dfrac{ln\left(\rho_{Y_0,Y}\sqrt{exp\left( \left( \dfrac{\sigma_{X_0}}{\xi} \right)^2\right)-1}\sqrt{exp\left(\left( \dfrac{\sigma_X}{\xi} \right)^2\right)-1}+1\right)}{\sigma_{X_0}\sigma_X}
\label{eq:rho_N2LN}
\end{equation} 
\rule{0.97\textwidth}{0.5pt}
\vspace{-0.5cm}
\end{figure*}
\normalsize
\vspace{-0.8cm}
\subsection{The SIR and Spectral efficiency distribution at a particular user location}
As derived in \eqref{eq:SIR_simplified}, \eqref{eq:Y} and \eqref{eq:X}, the downlink SIR ($\Gamma$) in dB domain, denoted by $\Gamma_{dB}$ can be expressed as
\begin{equation}
 \Gamma_{dB} = X_0 - X.
\end{equation}
The distribution of $\Gamma_{dB}$ is normal with mean $\mu_{\Gamma_{dB}}$ and standard daviation $\sigma_{\Gamma_{dB}}$ given by: 
\begin{equation} 
\nonumber
 \mu_{\Gamma_{dB}} = \mu_{X_0}  - \mu_X, ~~ \sigma_{\Gamma_{dB}} = \sqrt{\sigma_{X_0}^2 + \sigma_{X}^2 -2\rho_{X_0,X}\sigma_{X_0}\sigma_{X}}.
\end{equation}
In linear domain, the SIR ($\Gamma$) is LN distributed whose distribution is
\begin{equation}
f_\Gamma(\gamma) = \dfrac{\xi}{\sqrt{2\pi}\sigma_{\Gamma_{dB}}\gamma}exp\left[-\dfrac{\left(\xi ln(\gamma)- \mu_{\Gamma_{dB}}\right)^2 }{2\sigma_{\Gamma_{dB}}^2} \right].
\label{eq:f_gamma}
\end{equation}
It has been assumed that the UE situated at any particular specified location use suitable modulation and coding scheme in order to achieve Shannon information limit for its instantaneous SIR ($\Gamma$). Under such assumption, the SE ($R_u$) at that particular user location can be expressed as $R_u =  \text{ln}(1+ \Gamma)$ nats/s/Hz. The distribution of $R_u$ can be obtained from the distribution of SIR using change of variables 
\small{
\begin{equation}
\nonumber
\begin{aligned}
 & f_{R_u}(r)= f_{\Gamma}(x).\Big|\frac{\partial \Gamma}{\partial R_u}\Big|_{R_u = r}=\\
&\frac{\exp(r)}{\sqrt{2\pi\sigma_{\Gamma_{dB}}^2}(\text{exp}(r)-1)} \exp\left[-\frac{(\text{ln}\left(\exp(r)-1)-\mu_{\Gamma_{dB}}\right)^2}{2\sigma_{\Gamma_{dB}}^2}\right], \\ & ~~~~~~~~~~~~~~~~~~~~~~~~~~~~~~~~~~~~~~~~~~~~~~~~~~~~~~~~~~~~~~~~ 0 \leq r \leq \infty. \ \
\end{aligned}
\label{equn:RateDensityFun}
\end{equation}}
\normalsize
Subsequently, the downlink mean ($\overline{R}_u$) and \textit{``$\alpha$-percentile''} outage ($R_{u_{\text{outage}}}$) downlink spectral efficiency at that particular user location is given by:
\begin{equation}
\begin{aligned}
  \overline{R}_u = \int_0^{\infty} r f_{R_u}(r) {\rm d} r,~~ \int_{0}^{R_{u_{\text{outage}}}} f_{R_u}(r) {\rm d}r = \alpha. 
\end{aligned}
\label{eq:MeanOutageSE}
\end{equation}
\section{Numerical Results and Discussions}\label{sec:NumericalResults} 
\begin{table}[h]
\caption{\small{Key system parameters}}
\centering
\begin{tabular}{|m{4.5cm}|m{3cm}|}\hline
\textbf{Parameter} & \textbf{Value}\\ \hline
\multirow{2}{*}{Network topology} & Indoor office \cite[Fig. 1]{maccartney_2015}  \\ \cline{2-2}
&  Hexagonal cells with ISD = $500m$ for UMa \cite{3gpp36873}. \\ \hline
\multirow{2}{*}{Carrier frequency ($f_c$)} & $28$ GHz (mmWave)  \\ \cline{2-2}
&  $2$ GHz (sub 6 GHz) \\ \hline
\multirow{2}{*}{PLE  in mmWave office layout \cite{maccartney_2015}} & $1.1$ (LOS)  \\ \cline{2-2}
&  $2.7$ (NLOS) \\ \hline
\multirow{2}{*}{PLE in sub-6 GHz office layout \cite{3gpp36873}} & $1.69$ (LOS)  \\ \cline{2-2}
&  $4.33$ (NLOS) \\ \hline
\multirow{2}{*}{$\sigma_\chi$ in mmWave office layout \cite{maccartney_2015}} & $1.8$ dB (LOS)  \\ \cline{2-2}
&  $9.6$ dB (NLOS) \\ \hline
\multirow{2}{*}{$\sigma_\chi$ in sub-6 GHz office layout \cite{3gpp36873}} & $3$ dB (LOS)  \\ \cline{2-2}
&  $4$ dB (NLOS) \\ \hline
$\sigma_\chi$ in sub-6 GHz UMa \cite{3gpp36873} & $6$ dB \\
\hline
\multirow{2}{*}{Shape factor of Gamma distribution}  & 5 in LOS (assumed)\\ \cline{2-2}
 & 1 (NLOS) \\ \hline
 Co-efficient of correlation between  & $0.5$\\
 shadowing from APs \cite{3gpp36814} & \\
\hline 
\end{tabular}
\label{table_SystemParameter}
\vspace{-0.75cm}
\end{table}
In this section we show the utility of the approximation method presented in the work by matching them with simulation results in indoor office environment operating in mmWave as well as sub-6 GHz frequency band. For office environment, we follow the same AP and UE numbers as given in  \cite[Fig. 1]{maccartney_2015}. Out of $33$ UE locations illustrated in  \cite[Fig. 1]{maccartney_2015}, due to space limitation we show the downlink performance at two different locations only, marked as UE location $1$ and $14$.
We also use the approximation framework in a regular 7-cell hexagonal cellular network representing urban macro (UMa) scenario \cite{3gpp36873}, with serving base station located at the origin (0,0). In that case, the simulated and approximated downlink SIR experienced by a UE residing in \textit{cell center} (location coordinates: $(25, 0)$ (in polar coordinates, distance is in meters)) as well as in \textit{cell edge} (location coordinates: $(225, 0)$) have been compared.  
The important system parameters are summarized in Table \ref{table_SystemParameter}.\\
 \begin{figure*}
\centering
\begin{minipage}[b]{.3\textwidth}
\includegraphics[width= \linewidth]{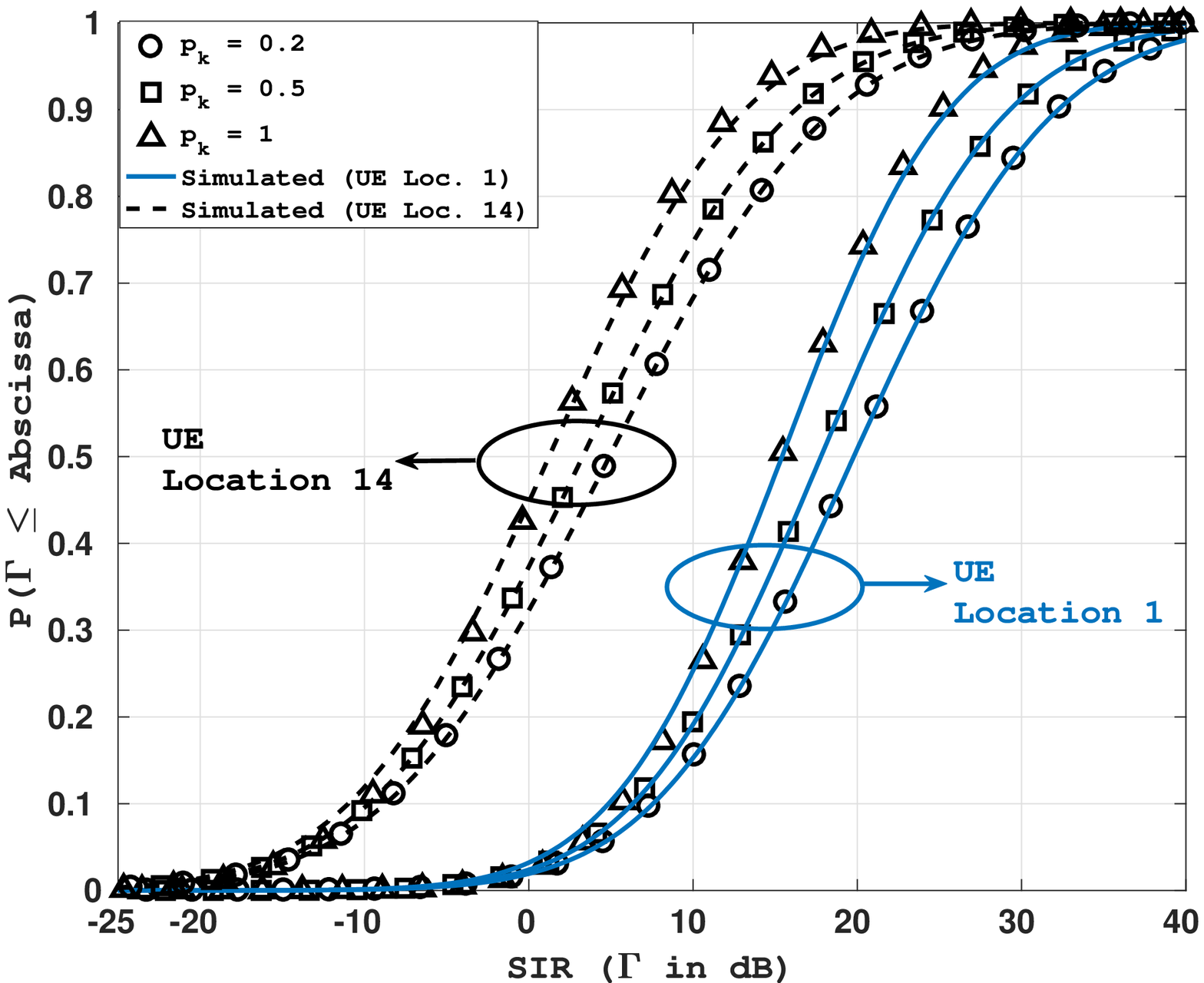}
\subcaption{\small{Office floor in mmWave band}}\label{fig:mmw_office_cdf}
\end{minipage}\qquad
\begin{minipage}[b]{.3\textwidth}
\includegraphics[width= \linewidth]{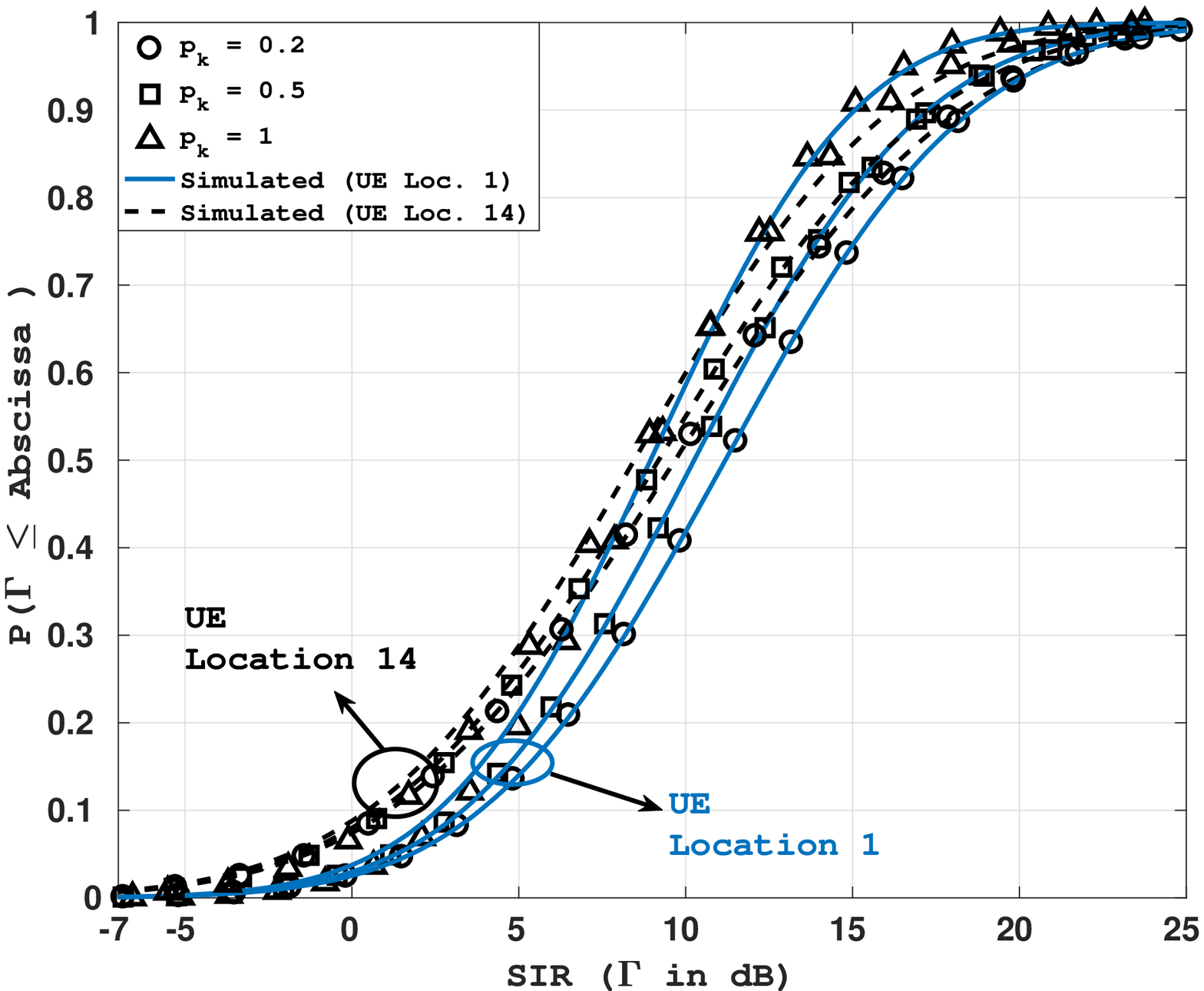}
\subcaption{Office floor in sub-6 GHz band}\label{fig:sub6_office_cdf}
\end{minipage}
\begin{minipage}[b]{.3\textwidth}
\includegraphics[width= \linewidth]{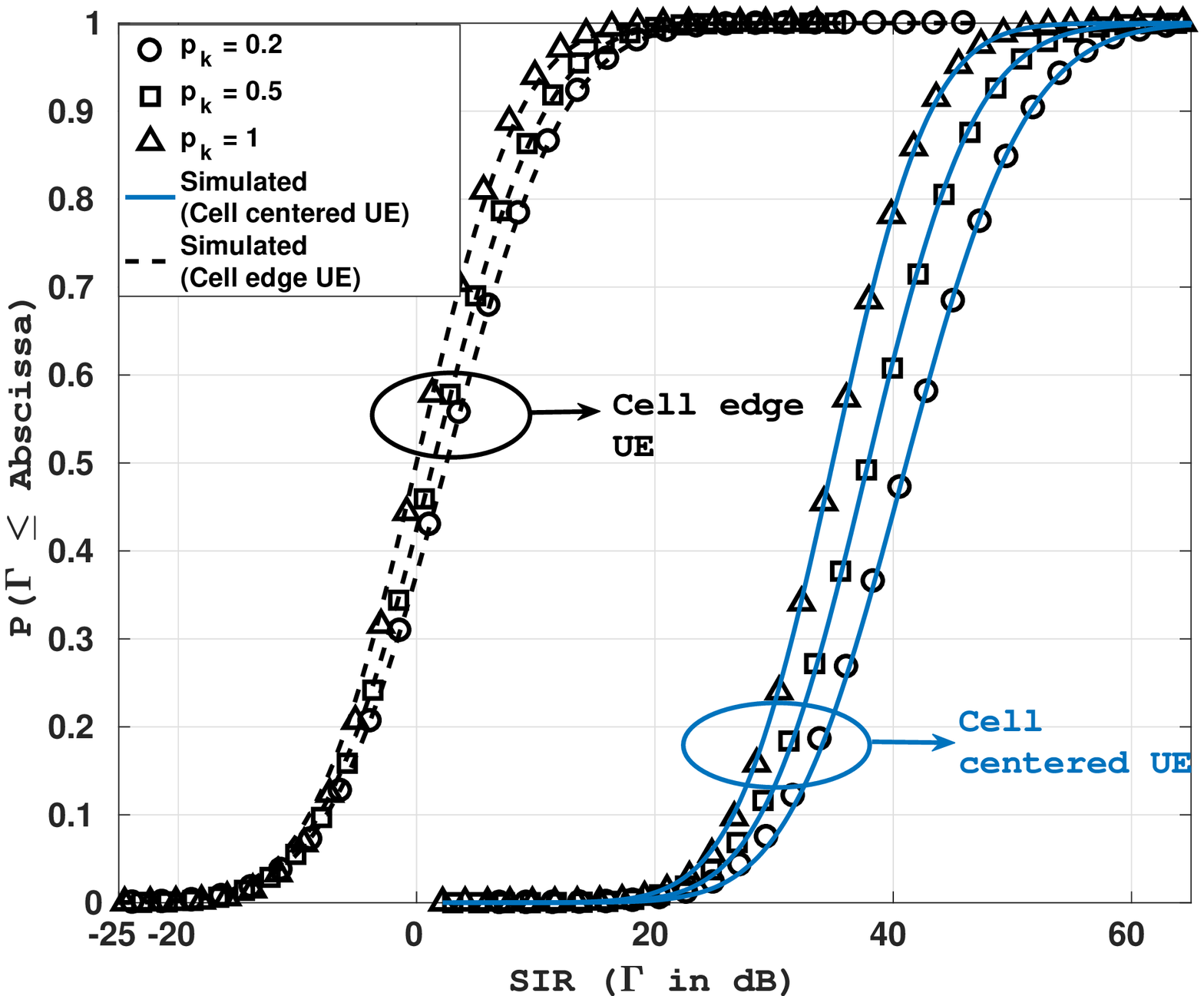}
\subcaption{Hexagonal UMa in sub-6 GHz band}\label{fig:sub6_UMa_cdf}
\end{minipage}
\caption{\small{CDF of SIRs ($\Gamma$, as in \eqref{eq:SIR_simplified}) for specific UE locations in indoor office and hexagonal network under mmWave and sub-6 GHz. Lines (both solid and dashed) represent the simulation results and markers represent the approximation results. }}\label{fig:cdf_sir}
\vspace{-0.5cm}
\end{figure*}
\begin{figure}
\vspace{-0.3cm}
   \centering
   \includegraphics[height=0.58\linewidth,width= 0.75\linewidth]{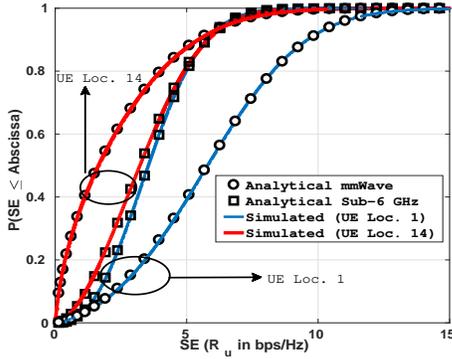}
   \caption{\small{CDF of SE for specific UE locations in indoor office under mmWave and sub-6 GHz ($p_k = 0.5$). }}
   \label{fig:mmw_sub6_office_cdf_SE}
   \vspace{-0.6cm}
\end{figure}
\begin{figure}
   \centering
   \includegraphics[height=0.58\linewidth,width= 0.75\linewidth]{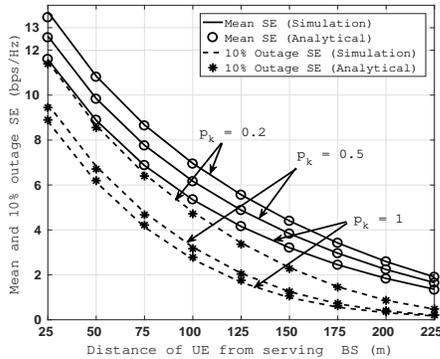}
   \caption{\small{Variation of mean and outage SE with respect to BS-UE distance in UMa}}
   \label{fig:varied_ue-location}
   \vspace{-0.7cm}
\end{figure}
\indent For LOS links, no typical value of $m$ has been provided in measurement campaigns. In this work we consider $m$ = 5 for LOS links, however, the framework is generic to accommodate any other value of $m$ as well. 
The modified values of $\mathscr{L}(d)$ (as in \eqref{eq:PL_LOS_NLOS_mmW}) and $\sigma_\chi$ to accommodate the small scale fading, are provided in Table \ref{tb:PLShadow}.
\begin{table}
\caption{\small{Modifications for $\mathscr{L}$ and $\sigma_\chi$ due to small scale fading.}}
\label{tb:PLShadow}
\centering
\begin{tabular}{|c|c|c|c|c|c|} 
\hline
Environment & Link   & $\sigma_\chi$  & $m$  & $\mathscr{L}(d) - \hat{\mathscr{L}}(d)$ &  $\hat{\sigma_\chi}$  \\
&  &  (dB) & &  (dB) &  (dB)\\
\hline
Indoor (28 GHz) \cite{maccartney_2015} & LOS  &  1.8 & 5 &  -0.45 & 2.72 \\
\cline{2-6}
& NLOS  &  9.6 & 1 & -2.51 & 11.1 \\
\hline
Indoor (2 GHz) \cite{3gpp36873} & LOS  &  3 & 5 &  -0.45 & 3.63 \\
\cline{2-6}
& NLOS  &  4 & 1 & -2.51 & 6.86 \\
\hline
UMa (2 GHz) \cite{3gpp36873} & NLOS  &  6 & 1 & -2.51 & 8.19 \\
\hline
\end{tabular}
\vspace{-0.5cm}
\end{table}
The statistical parameters of CCI ($\mu_X$ and $\sigma_X$) have been obtained by solving \eqref{eq:final} using standard \texttt{fsolve} function in MATLAB with $(s_1, s_2) = (1.0, 0.2)$, as provided in \cite{mehta_2007}.
In order to present a pessimistic estimate of the performance we assume that for any UE location, the nearest interferer AP (i.e. the second nearest among all APs) is having full activity, i.e. probability that it is switched on is $1$. The activity factor ($p_k$) of all other interferers are considered to vary among $0.2, 0.5$ and $1$ in order to represent low, moderate and full level of activity respectively.\\
\indent In Fig. \ref{fig:mmw_office_cdf} and \ref{fig:sub6_office_cdf}, the distributions of $\Gamma$ (in \eqref{eq:SIR_simplified}) at two different UE locations in indoor office layout have been shown for mmWave and sub-6 GHz bands respectively for a wide range of $p_k$. Along with that, the distributions of $\Gamma$ at cell center and cell-edge UE locations in hexagonal layout (UMa) has also been shown in Fig. \ref{fig:sub6_UMa_cdf}. The obtained close match between the simulated and approximated CDF curves shows the usability of \eqref{eq:f_gamma}. Furthermore, with increase in $p_k$, the aggregate interference power received at the UE locations increases, thus resulting in decrease in mean SIR, as shown in Fig. \ref{fig:cdf_sir}. \vspace{-0.2cm}
\begin{table}[h]
\centering
\caption{\small{Relative error between approximated and simulated distributions of SIR in terms of KLD and KSD for different scenarios in different frequency bands ($p_k = 0.5$)}}
\begin{minipage}[b]{.5\textwidth}
\centering
\resizebox{\columnwidth}{!}{
\begin{tabular}{|c|c|c|c|c|}\hline
\multirow{2}{*}{UE Location} & \multicolumn{2}{c|}{Sub-6 GHz}             & \multicolumn{2}{c|}{mmWave}                            \\  \cline{2-5}
                             & KLD                 & KSD           & KLD                  & KSD                    \\ \cline{1-5}
UE 1                         & $11\times 10^{-4}$  & $4.8\times 10^{-3}$  & $8.58\times 10^{-4}$ & $5\times 10^{-3}$  \\ \hline
UE 14                        & $5.71\times10^{-4}$ & $6.7\times 10^{-3}$  & $3.86\times 10^{-4}$ & $1.3\times 10^{-3}$  \\ \hline
\end{tabular}}
\subcaption{\small{Office scenario}}
\label{tab:kld_ksd_table_indoor}
\end{minipage}
\begin{minipage}[b]{.5\textwidth}
\centering
\begin{tabular}{|c|c|c|}\hline
UE Location   & KLD  & KSD  \\ \hline
Cell centered & $13\times 10^{-4}$ & $5.7\times 10^{-3}$ \\ \hline 
Cell edge     & $5\times 10^{-4}$ & $7.7\times 10^{-3}$ \\ \hline
\end{tabular}
\subcaption{\small{UMa scenario}}
\label{tab:kld_ksd_table_UMa}
\end{minipage}
 \label{tab:kld_ksd_table}
 \vspace{-0.8cm}
\end{table}
\begin{table}[h]
\caption{\small{Mean and 10$\%$ outage SEs (simulated and calculated, in bps/Hz) for specific UE locations in indoor office ($p_k = 0.5$).}}
\centering
\resizebox{\columnwidth}{!}{
\begin{tabular}{|c|c|c|c|c|c|}\hline
\multirow{2}{*}{Freq. band} & \multirow{2}{*}{UE Locations} & \multicolumn{2}{c|}{Simulation}                                        & \multicolumn{2}{c|}{Analytical}                                         \\ \cline{3-6}
                            &                               & Mean SE & \begin{tabular}[c]{@{}l@{}}$10\%$ \\ outage SE\end{tabular} & Mean SE & \begin{tabular}[c]{@{}l@{}}$10 \%$ \\ outage SE\end{tabular} \\ \hline
\multirow{2}{*}{mmWave}     & UE 1                          & 5.91    & 2.28                                                        & 5.90    & 2.27                                                         \\
                            & UE 14                         & 2.29    & 0.146                                                        & 2.28    & 0.151                                                         \\ \hline
\multirow{2}{*}{Sub-6 GHz}  & UE 1                          & 3.65    & 1.655                                                        & 3.69    & 1.631                                                         \\
                            & UE 14                         & 3.36    & 1.13                                                        & 3.35    & 1.142   \\ \hline                                                     
\end{tabular}}
    \label{tab:KPI_results}
    \vspace{-0.7cm}
\end{table}
From the floor layout presented in \cite[Fig. 1]{maccartney_2015}, it can be seen that in UE location 1, the serving AP is in LOS condition, whereas in UE location 14, the serving AP is in NLOS position. In mmWave scenario, the propagation loss is significantly higher in NLOS condition compared to LOS condition, thus resulting in significant decrease in mean received SIR in UE location 14 compared to UE location 1, as inferred from Fig. \ref{fig:mmw_office_cdf}. On the other hand, in sub-6 GHz, the propagation loss in NLOS condition is marginally higher than LOS condition, which results in marginal decrease in mean received SIR in UE location 14 compared to UE location 1, as inferred from Fig. \ref{fig:sub6_office_cdf}.\\
\indent The fitness of the presented approximation method is highlighted in Table \ref{tab:kld_ksd_table} where the relative error between the simulated and approximated CDF of SIR has been quantified in terms of two statistical metrics namely Kullback–Leibler divergence (KLD) and Kolmogorov–Smirnov distance (KSD) for both indoor office and urban macro scenario (in Table \ref{tab:kld_ksd_table_indoor} and \ref{tab:kld_ksd_table_UMa} respectively) for moderate interferer activity ($p_k = 0.5$). Extremely low values of such metrics (in the order of $10^{-4}$ for KLD and of $10^{-3}$ for KSD) shows the validity of the proposed method across various scenarios in different UE locations\footnote{\label{note_2}The results are similar for low and full level of interferer activity as well and thus omitted for brevity.}.\\
\indent The distributions along with the mean and $10\%$ outage SE at two different UE locations in indoor office layout have been shown for $p_k = 0.5$ in Fig. \ref{fig:mmw_sub6_office_cdf_SE} and Table \ref{tab:KPI_results} respectively. The simulated and approximated results are found to be reasonably close, which in turn shows the usability of \eqref{eq:MeanOutageSE} for indoor office network layout in sub-6 GHz as well as in mmWave bands\textsuperscript{\ref{note_2}}. One important observations is that, when the UE is under LOS coverage (UE loc. 1), values of mean and  outage SE are higher at mmWave band than at sub-6 GHz band. But, where the UE in under NLOS coverage (UE loc. 14), the mean and outage SE is higher at sub-6 GHz than at mmWave band. This further establishes the superiority of mmWave communication in LOS conditions.   \\
\indent Finally, in Fig. \ref{fig:varied_ue-location}, the mean and outage SE performances have been shown in hexagonal UMa scenario by varying the distance of UE from the serving BS (situated at origin) from 25 m to 225 m across the horizontal axis in order to capture the performance throughout the cell i.e. from cell center to cell edge. A quite close match between the simulated and approximated results indicates the usefulness of the presented framework. With increase in UE-BS separation, the UE moves from cell center to cell edge, thus resulting in a decrease in mean and $10\%$ outage SE. With increase in the interferer's activity level from low ($p_k = 0.2$) to moderate ($p_k = 0.5$) and then to full ($p_k = 1$), the aggregate interference power increases, resulting in decrease in mean and $10\%$ outage SE. 
\vspace{-0.4cm}
\section{Conclusions}
\label{sec:conclusions}
In this work we have derived the SIR distribution and hence SE results considering the holistic effect of LOS/NLOS SSF, LN distributed correlated desired and interferer shadowing along with interferer activity. The analytical expressions yield values of SIR and SE which very closely match with simulation results in both sub-6 as well as mmWave bands in indoor and outdoor scenarios for low, moderate as well as full activity level of neighboring interferers. The presented work covers the existing gap in literature which do not consider above-mentioned practical effects which are usually suggested in standard evaluation methods. Thus, this work can be applicable in analytical performance evaluation of future wireless communication systems in realistic scenario.    
\vspace{-0.45cm}
\bibliographystyle{IEEEtran}
\bibliography{ref_list} 
\end{document}